# Effect of Fe-substitution on the structure and magnetism of single crystals $Mn_{2-x}Fe_xBO_4$

M.S. Platunov[1,2], N.V. Kazak*[1], Yu.V. Knyazev[1], L.N. Bezmaternykh[1], E.M. Moshkina[1], A.L. Trigub[3], A.A. Veligzhanin[3], Y.V. Zubavichus[3], L.A. Solovyov[4], D.A. Velikanov[1], S.G. Ovchinnikov[1]

[1]*Kirensky Institute of Physics, Federal Research Center KSC SB RAS, 660036 Krasnoyarsk, Russia*
[2]*ESRF-The European Synchrotron, F-38043 Grenoble 9, France*
[3]*National Research Centre "Kurchatov Institute", 123182 Moscow, Russia*
[4]*Institute of Chemistry and Chemical Technology, Federal Research Center KSC SB RAS, 660049 Krasnoyarsk, Russia*

**Abstract** Single crystalline $Mn_{2-x}Fe_xBO_4$ with $x$ = 0.3, 0.5, 0.7 grown by the flux method have been studied by means of X-ray diffraction and X-ray absorption spectroscopy at both Mn and Fe $K$ edges. The compounds were found to crystallize in an orthorhombic warwickite structure (sp. gr. $P_{nam}$). The lattice parameters change linearly with $x$ thus obeying the Vegard's law. The $Fe^{3+}$ substitution for $Mn^{3+}$ has been deduced from the X-ray absorption near-edge structure (XANES) spectra. Two energy positions of the absorption edges have been observed in Mn $K$-edge XANES spectra indicating the presence of manganese in two different oxidation states. Extended X-ray absorption fine structure (EXAFS) analysis has shown the reduction of local structural distortions upon Fe substitution. The magnetization data have revealed a spin-glass transition at $T_{SG}$= 11, 14 and 18 K for $x$= 0.3, 0.5 and 0.7, respectively.

## 1. Introduction

The oxyborates with warwickite structure $M^{2+}M'^{3+}BO_4$ ($M^{2+}$ and $M'^{3+}$ are Mg, Sc or 3d ion) are attracting considerable interest as a magnetic frustrated systems with quasi low-dimensionality effects [1, 2]. 1D magnetic behavior has been identified experimentally in $MgTiBO_4$ warwickite (S=1/2) which does not undergo magnetic ordering down to 1.4 K [3, 4]. Recently, the interest has grown due to the physical realization of quantum entanglement in these systems [5, 6]. In addition, two homometallic warwickites $Mn_2BO_4$ and $Fe_2BO_4$ described so far show a charge ordering [7-11] and a long-range magnetic order below $T_N$=26 and 155 K, respectively [8, 12, 13].

The synthesis of the single-crystalline warwickites is a problem of enhanced complexity. These compounds include both divalent and trivalent metal ions and demonstrate a high sensitivity of the physical properties to the little content changes. Moreover the complex crystal structure and quasi-low-dimensional arrangement of the structural units cause the strong growth anisotropy, which complicates the obtaining of the single crystal of suitable size for the physical experiment.

There are a number of a works devoted to preparation of the oxyborates with warwickite structure by different techniques. These techniques are the solid-state reaction method [1, 3, 4, 9, 13] flux technique [6, 7, 14, 15] and in some cases a hydothermal technique [16]. Solid-state reaction synthesis consists of the mixing of the initial crystal-forming oxides in a stoichiometric ratio $M^{2+}M'^{3+}BO_4$ (in some cases with the excess of the $H_3BO_3$ for compensation of the boron volatilization) and the annealing at high temperatures. As a result the polycrystalline samples with the size about $0.01*0.01*0.05$ mm$^3$ were obtained. The flux technique is often used, where the solvent is the boron oxide $B_2O_3$ or borax $Na_2B_4O_7$. However, the use of these solvents allows one to synthesize single crystals with a size of not more than 1 mm [17]. The most comprehensive and systematic study of the crystallogenesis and the structure of the prepared by different technique warwickite series has been carried out in the well-known work of *Capponi et.al.* [15]. Experimental investigation of the stability and the crystallization peculiarities of the warwickites in the boron fluxes ($H_3BO_3$) have been performed at the temperatures 900 – 1200°C and the pressures of 40-80 kBar using carbonates and fluorides of the alkaline, transition and rare-earth metals. This has allowed studying the influence of the different types of the cations to the crystallization process. As a

———————————————————————
*Corresponding author:*
Dr. Natalia Kazak
Kirensky Institute of Physics, FRC KSC SB RAS
Akademgorodok 50/38, Krasnoyarsk, 660036, Russia
Tel.: +7(391)249-45-56, Fax: +7(391)243-89-23
E-mail: nat@iph.krasn.ru

result the series of the warwickites with $M^{2+}$ = Mg, Ni, Co, Fe, Mn, Cd, Ca and $M'^{3+}$ = Al, Ga, Cr, Fe, Mn, V, Ti, Sc, In, Lu, Yb, Tm, Y, Dy have been obtained.

The special case of the warwickites synthesis problem is the homometallic warwickites. For synthesis of the polycrystalline samples of $Fe_2BO_4$ the solid-state reaction technique was used, and metallic iron Fe was added to the initial mixture to obtain a divalent cation [9]. This circumstance limits the range of the synthesis techniques, which could be used for preparation of this compound. The synthesis of the $Mn_2BO_4$ has been performed as by the solid-state reaction technique [7, 9, 13], as by the flux method [7, 8]. For solid-state reaction method the initial components were manganese carbonate $Mn^{2+}CO_3$ and boric acid $H_3BO_3$ with annealing at 700°C and 800°C (I) [9], or manganese oxide $Mn_2O_3$ and boron oxide $B_2O_3$ with annealing at 800°C (II) [7], or $Mn(NO_3)_2 \cdot 4H_2O$ and boric acid $H_3BO_3$ with annealing at 700°C (III) [7, 13]. During the synthesis process it was supposed the ratio between the $Mn^{3+}$ and $Mn^{2+}$ cations corresponds to the ratio between these components in the initial system even at the high temperatures, at that the chemical reaction of the $Mn_2O_3 \rightarrow Mn_3O_4$ with loss of the oxygen at the temperatures 950-1100°C wasn't taken into consideration. The solid state reaction method was found to give rise to difficulties in the preparing of pure samples. The authors of Ref's [9, 13] marked the presence of magnetic impurities in the form of $Mn_2O_3$ and $Mn_3O_4$ oxides. The pure oxyborate phase has been obtained only in the case of (III) [7].

Recently, manganese oxyborate $Mn_2BO_4$ has been grown by the flux method using the system of $Bi_2Mo_3O_{12}$–$B_2O_3$–$Na_2CO_3$-$Mn_2O_3$ [8]. The X-ray diffraction has confirmed the formation of the warwickite structure without any secondary phases. The single crystalline samples were in the form of black needles up to 12 mm long, and the cross sectional area was ~ 1.0*0.5 mm$^2$ that allowed performing the anisotropic properties study.

To the best of our knowledge, only two papers have been published so far reporting the crystal structure of $MnFeBO_4$ [15,18]. This compound has been synthesized by the flux method and the obtained samples have the size of 0.02*0.06*0.1 mm$^3$ due to the using of the high viscosity boron-based flux.

The aim of this study was to investigate conditions for the growth of $Mn_{2-x}Fe_xBO_4$ (0< $x$ <1) single crystals from the $Bi_2Mo_3O_{12}$–$B_2O_3$–$Na_2O$–$Mn_2O_3$–$Fe_2O_3$ flux system, that is one of the most successful methods for production of the single-crystalline samples of high quality and sufficient size. In present study the question about the oxidation state of the host and substitution ions has been investigated using element-specific XANES at the Mn and Fe $K$-edges. Additionally, the local environment of Mn and Fe atoms has been studied thoroughly using EXAFS. Preliminary study of magnetic properties has shown a low-temperature spin-glass state below $T_{SG}$=11, 14 and 18 K for $x$=0.3, 0.5 and 0.7, respectively.

## 2. Sample preparation and instrumental characterization

Single crystalline samples of $Mn_{2-x}Fe_xBO_4$ were grown by the flux method in the system (100 - $n$) mass% ($Bi_2Mo_3O_{12}$ + $p \cdot B_2O_3$ + 0.7·$Na_2O$) + $n$ mass% ($Mn_2O_3$ + $q \cdot Fe_2O_3$), where $n$ is the crystal-forming oxide concentration corresponding to the $Mn_{2-x}Fe_xBO_4$ stoichiometry, $p$ and $q$ are parameters of the crystal-forming oxide ratio to the matrix. Fluxes with different $p$, $q$, and $n$ and a total mass of 83-90 g were prepared by sequential melting of high purity oxides $Bi_2Mo_3O_{12}$, $B_2O_3$, ($Mn_2O_3$ and $Fe_2O_3$), and $Na_2CO_3$ in a 100 cm$^3$ platinum crucible. The main crystallization parameters are presented in Table 1. The flux homogenization was completed within 3 h at $T$=1100 °C, then the melt was subjected to two-stage cooling: fast cooling to $T$=($T_{sat}$-10) °C followed by slow cooling at a rate of 4 °C/day for three days. The melt was then poured out. Warwickite was the only crystallizable phase at temperature ~ 40 °C. The grown single crystals were subjected to etching in 20% nitric acid. As a result, single crystals in the form of black prisms with a typical size of 0.4 × 0.2 × 8.0 mm$^3$ strongly extended along the $c$-axis were obtained (inset to Fig. 1).

Room temperature X-ray powder diffraction (PXRD) data for $Mn_{2-x}Fe_xBO_4$ were collected on a PANalytical X'Pert PRO diffractometer equipped with a solid-state detector PIXcel using Co-$K\alpha$ radiation over the $2\theta$ range 10–90° (Fig. 1). The crystal lattice parameters were refined using the Rietveld method [19] with the full-profile derivative difference minimization [20].

**Table 1**
The crystallization parameters for $Mn_{2-x}Fe_xBO_4$ warwickites.

| Target stoichiometry | $p$ | $q$ | $n$, % | $T_{sat}$, °C |
|---|---|---|---|---|
| $Mn_{1.75}Fe_{0.25}BO_4$ | 2.21 | 0.08 | 24.6 | 925 |
| $Mn_{1.5}Fe_{0.5}BO_4$ | 2.61 | 0.16 | 25.5 | 880 |
| $Mn_{1.25}Fe_{0.75}BO_4$ | 2.87 | 0.24 | 26.5 | 875 |

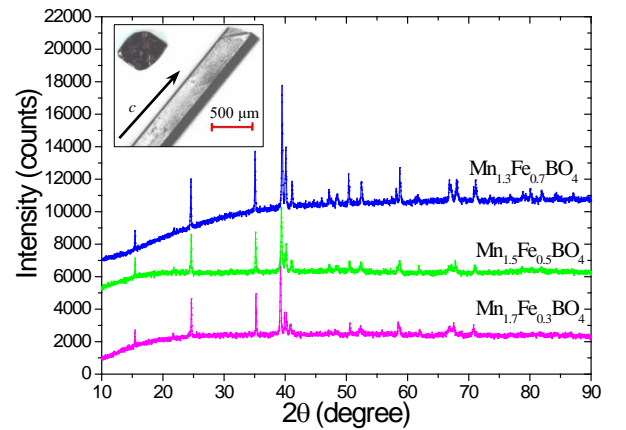

**Fig. 1.** PXRD patterns of $Mn_{2-x}Fe_xBO_4$ samples. The inset: photo of single crystal of $Mn_{2-x}Fe_xBO_4$ warwickite in $ab$-plane and along $c$-axis.

XANES and EXAFS spectra at the Mn and Fe $K$-edges were recorded at room temperature in the transmission mode using a SHI closed-cycle helium refrigerator (Japan) at the Structural Materials Science beamline of the Kurchatov Synchrotron Radiation Source (National Research Center "Kurchatov Institute", Moscow) [21]. For the selection of the primary beam photon energy, a Si(111) channel-cut monochromator was employed, which provided an energy resolution $\Delta E/E \sim 2\times10^{-4}$. Incident and transmitted intensities were recorded using two ionization chambers filled with appropriate $N_2$/Ar mixtures to provide 20% and 80% absorption. The energies were calibrated against a sharp pre-edge feature of $KMnO_4$ (Mn $K$-edge) and using Fe metal foil (Fe $K$-edge). The EXAFS spectra were collected using optimized scan parameters of the beamline software. The $\Delta E$ scanning step in the XANES region was about 0.45 eV, and scanning in the EXAFS region was carried out at a constant step on the photoelectron wave number scale with $\Delta k = 0.05$ Å$^{-1}$ that corresponds to the energy step of the order of 1.5 eV. The signal integration time was 4 s per point. Single-crystalline samples were ground to fine powders and then spread uniformly onto a thin adhesive Kapton film which was folded several times to provide an absorption edge jump around unity.

The EXAFS spectra $\mu(E)$ were normalized to a unit edge jump and the isolated atom absorption coefficient $\mu_0(E)$ was extracted by fitting a cubic-spline-function versus the experimental data. After subtraction of the smooth atomic background, the conversion from photon energy $E$ to photoelectron wave number $k$ scale was performed.

Crystallographic parameters were used as a starting structural model. The $k^3$-weighted EXAFS function $\chi(k)$ was calculated in the intervals $k = 2 – 11.5$ Å$^{-1}$ using a Hanning window (local order peaks were clearly distinguishable against background up to 7Å). The EXAFS structural analysis was performed using theoretical phases and amplitudes as calculated by the FEFF8 package [22], and fits to the experimental data were carried out in the $R$-space with the IFFEFIT package [23].

The $dc$ magnetization measurements were carried out using a SQUID magnetometer [24] for parallel orientation of applied field respective the $c$-axis in magnetic field of 250 Oe. The temperature interval is 2-270 K.

## 3. Experimental results
### 3.1. X-ray powder diffraction data

The X-ray powder diffraction has revealed the pure warwickite-type structure, the space group *Pnam* for all samples $Mn_{2-x}Fe_xBO_4$. The warwickite structure is built by low-dimensional ribbons of four edge-sharing oxygen octahedra running along the $c$-axis (Figure 2). Divalent and trivalent metal ions randomly occupy two nonequivalent crystallographic sites both belonging to general 4$c$ Wyckoff positions (M1 and M2). There are four formula units per unit cell. The room temperature lattice parameters of $Mn_{2-x}Fe_xBO_4$ are varied linearly with Fe content obeying the Vegard's law (Figure 3, Table SM1 [25]). The $b$ parameter increases, while $a$ and $c$ decrease.

The Mn/Fe metal ratios in the samples studied were independently controlled by experimental absorption edge jumps in X-ray absorption spectra (Figure 4) [26]. The calculated Fe stiochiometric coefficients were found to be $x = 0.34$, 0.53 and 0.72. These compositions are in a good agreement with those predicted by the synthesis technique. Hereinafter, we shall use chemical formulas $Mn_{1.7}Fe_{0.3}BO_4$, $Mn_{1.5}Fe_{0.5}BO_4$ and $Mn_{1.3}Fe_{0.7}BO_4$ for three samples for the sake of simplicity.

### 3.2. XANES spectra

Normalized XANES spectra at the Fe $K$-edge and their first derivatives for the three warwickite samples $Mn_{2-x}Fe_xBO_4$ and reference binary oxides FeO and α-$Fe_2O_3$ at room temperature are shown in Figure 5. All samples' spectra demonstrate a prominent similarity. A weak pre-edge absorption feature at 7115 eV arises from the electronic transition from 1$s$ to bound Fe 3$d$ states that are allowed by symmetry due to mixing with the oxygen 2$p$ band. The weak intensity of the pre-edge peak indicates a slightly distorted six-coordinated octahedral environment around the Fe atom as it is expected for the warwickite structure. The main edge position ($E_0^{Fe}$ = 7129 eV) observed for the $Mn_{2-x}Fe_xBO_4$ warwickites is the same as that for $Fe_2O_3$, clearly indicating the $Fe^{3+}$ state.

The background-subtracted pre-edge structure at the Fe $K$-edge for $Mn_{1.3}Fe_{0.7}BO_4$ is presented in Figure 6. A broader doublet structure with centroid at 7116 eV assigned to the transition of the 1$s$ electron to the 3$d$ crystal-field-split states is observed. We have fitted the double-peak pre-edge structure with a sum of two Gaussians, centered at 7115.8 eV (A1) and 7117.1eV (A2). The expanded pre-edge structure is shown at bottom panel of Figure 6. For the high-spin $Fe^{3+}$ ion ($^5A_{1g}$ ground state), two electronic transitions are expected: the $^5T_{2g}$ and $^5E$ with the energy splitting of the magnitude of $10Dq$. The separation of these two peaks was found to be $10Dq = 1.30 \pm 0.05$ eV. We noted that this parameter is in good agreement with those reported for $FeBO_3$ $10Dq = 1.4$ eV from RIXS data [27].

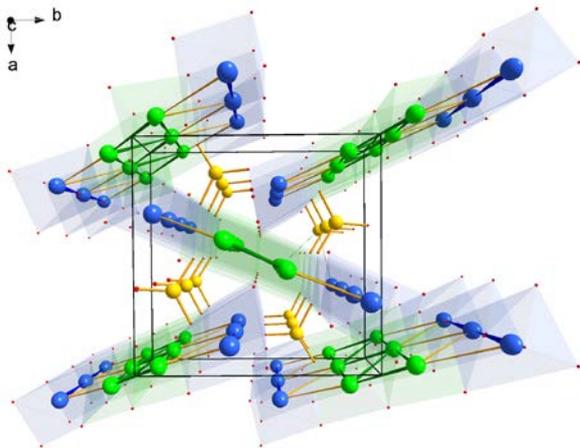

**Fig. 2.** Crystal structure of $M_2BO_4$ warwickites projected along [001]. The two distinct crystallographic sites M1 and M2, octahedrally coordinated by oxygen atoms, are shown as green and blue polyhedra, respectively. The boron atoms are represented by yellow. The low-dimensional ribbons running along $c$-axis are visible. The intra-ribbon couplings give rise to the geometric frustration.

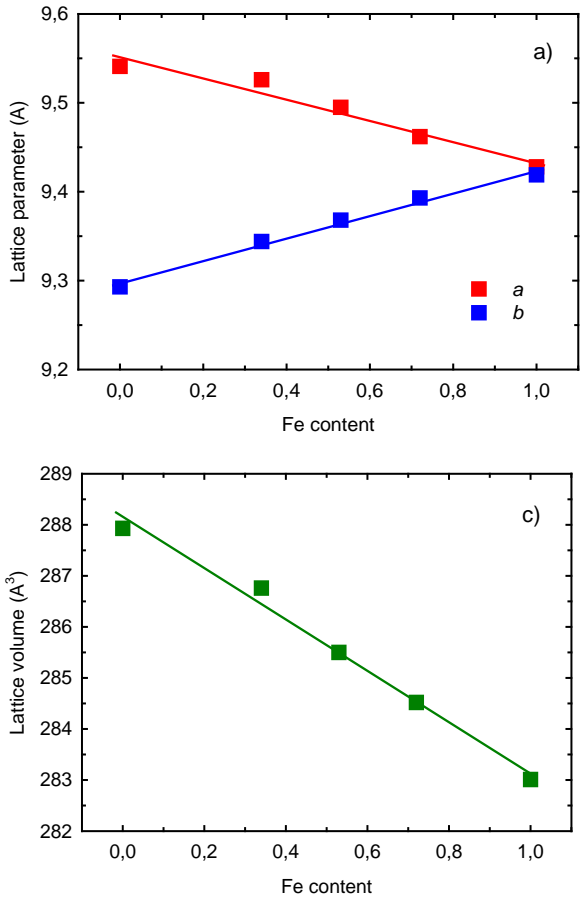
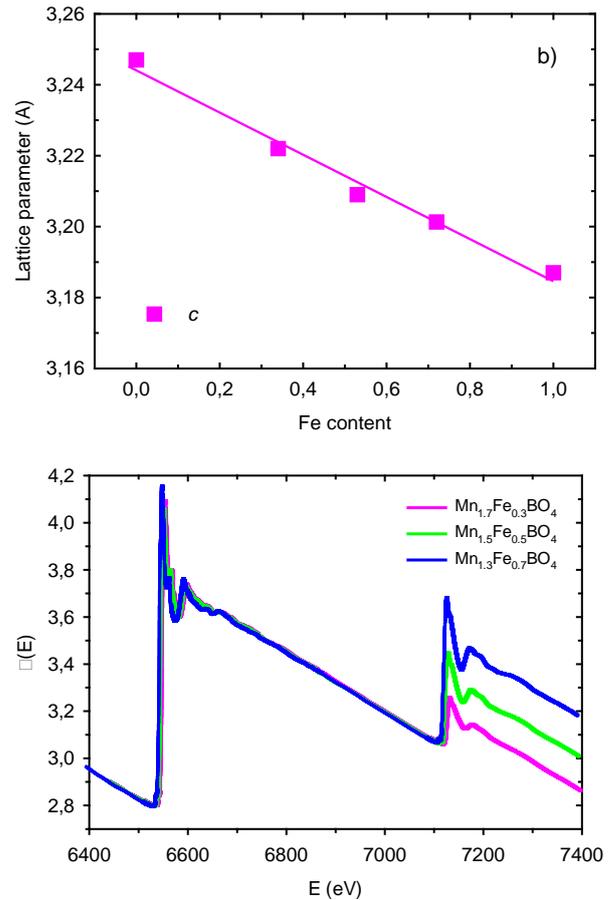

**Fig. 3.** The variation of lattice parameters with $x$ in $Mn_{2-x}Fe_xBO_4$ warwickites. a) $a$ and $b$ parameters, b) $c$ parameter, c) $V$ lattice volume. The straight lines are drawn for clarity. The lattice parameters for $MnFeBO_4$ are taken from Ref. [15].

**Fig. 4.** XAS data of the Mn and Fe $K$ absorption edges of the $Mn_{2-x}Fe_xBO_4$ samples. The increase in a Fe $K$ edge jump correlates with Fe content.

Figure 7a presents the normalized Mn $K$-edge XANES spectra and their first derivatives for the $Mn_{2-x}Fe_xBO_4$ warwickites obtained at room temperature. They are compared with those of $MnB_2O_4$ and $Mn_2O_3$ as references of the $Mn^{2+}$ and $Mn^{3+}$ states. XANES spectrum of $Mg_{0.7}Mn_{1.3}BO_4$ sample has been also measured as an example of manganese mixed-valence state in the warwickite structure. The following main features can be marked: i) the weak pre-edge peak (*A*) appear at ~6541eV consistent with a somewhat distorted centrosymmetric octahedral environment around the Mn atoms. ii) All curves show a clear shoulder (*B*) at the onset of the absorption edge at ~6550 eV, which is transformed into a distinct step for $Mn_{1.3}Fe_{0.7}BO_4$ and $Mg_{0.7}Mn_{1.3}BO_4$ samples. iii) The sample's derivatives spectra demonstrate two distinct maxima at ~ 6548 eV and ~6552 eV. The positions of the maxima are independent of the stoichiometry (Figure 7b) and are presumably assigned to the ionization thresholds of $Mn^{2+}$ and $Mn^{3+}$ ions in the warwickite structure. iv) The intensity of main peak (*C*) at ~ 6553eV enhances with increasing $x$. This effect can be explained by the difference in the degree of 3$d$- 4$p$ orbital mixing arising from the variation of Mn local structure with the Fe substitution. The monoclinic $Mn_2BO_4$ is characterized by a highly distorted local structure around $Mn^{3+}$ due to the JT effect. The Fe substitution gives rise to the orthorhombic crystal structure with a lower degree of local distortions around the manganese atoms.

In general, the Mn valence state in substituted warwickite $Mn^{2+}_{1-y}M^{2+}_y Mn^{3+}_{1-x}M^{3+}_x BO_5$ is determined by the expression

$$v^{Mn} = \frac{2 \cdot (1-y) + 3 \cdot (1-x)}{(1-y) + (1-x)}$$

where $x$, $y$ are the concentration of $M^{3+}$ and $M^{2+}$ ions, respectively. For $Mn_{2-x}Fe_xBO_4$ system, where $y=0$ and $x$ is 0.0, 0.3, 0.5 and 0.7, the $v^{Mn}$ are expected to be ~2.5, 2.4, 2.32 and 2.22 respectively.

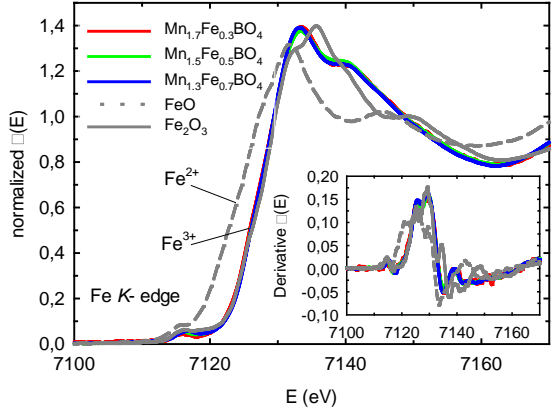
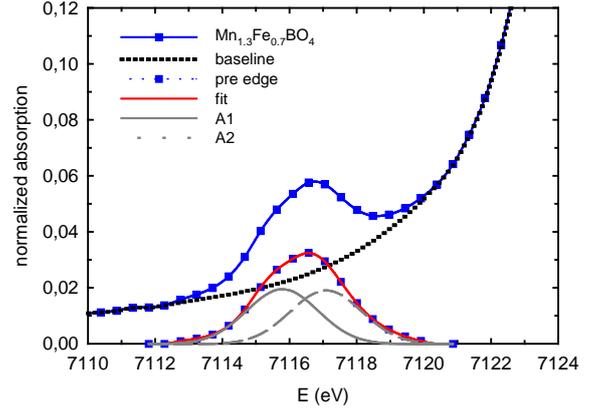

**Fig. 5.** Normalized Fe *K*-edge XANES spectra of FeO, $Fe_2O_3$ and $Mn_{2-x}Fe_xBO_4$ warwickites at room temperature. The inset shows first derivatives of the spectra.

**Fig. 6.** Fit of Fe *K*-edge pre-edge region for $Mn_{1.3}Fe_{0.7}BO_4$ using two Gaussian peak functions (A1 and A2).

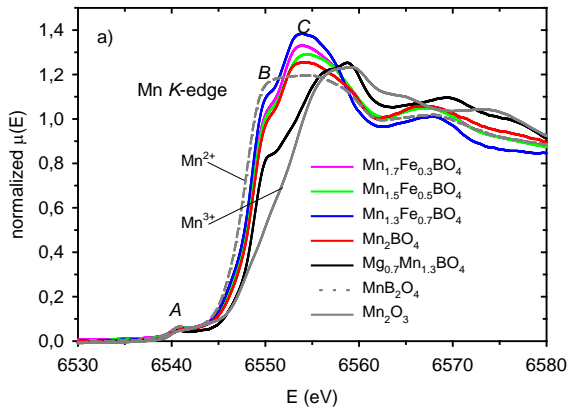
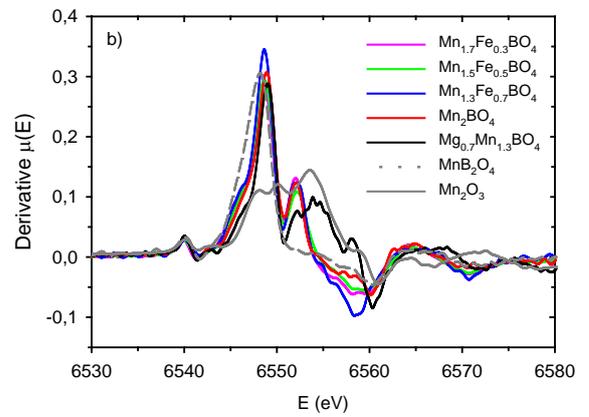

**Fig. 7.** a) Normalized Mn*K*-edge XANES spectra of $MnB_2O_4$, $Mn_2O_3$ and $Mn_{2-x}Fe_xBO_4$, $Mg_{0.7}Mn_{1.3}BO_4$ warwickites at room temperature. b) First derivatives of the spectra (same colors).

For $Mg_{0.7}Mn_{1.3}BO_4$ sample where $y=0.7$ and $x=0$ the $v^{Mn}$ reaches the value of 2.78. The Mn-Fe and Mg-Mn samples demonstrate a sharp rising of absorption feature at $E_0=6548$ eV. The growth of the *B*-peak intensity from $Mg_{0.7}Mn_{1.3}BO_4$ to $Mn_{1.3}Fe_{0.7}BO_4$ reflects the increase of the $Mn^{2+}$ content (22% in the Mg-Mn and 78% in Mn-Fe). An increase in this transition intensity effectively shifts the onset of the absorption spectrum to the left in the case of $Mn_{1.3}Fe_{0.7}BO_4$. Despite the fact that the effective manganese valence changes by ~0.6 on going from $Mg_{0.7}Mn_{1.3}BO_4$ to $Mn_{1.3}Fe_{0.7}BO_4$ samples the respective shift in the positions of the first derivative maximum does not exceed 0.3 eV. This implies that the mixing between the $Mn^{2+}$ and $Mn^{3+}$ electronic states in Mn-based warwickites is weak and thus the compounds should be referred to as mixed-valence rather than intermediate-valence ones.

*3.3. EXAFS*

Figures 8a and 8b show the moduli of Fourier transforms (FT) of EXAFS functions at Fe and Mn *K*-edges, respectively. The peaks within the distance range 1.0-3.5 Å correspond to the first Fe(Mn)-O and second Fe(Mn)-Fe, Fe(Mn)-Mn and Fe(Mn)-O coordination shells. The positions and shapes of the first FT peak in Fe K-edge EXAFS spectra are essentially similar for all samples. This indicates a rather symmetrical environment around the $Fe^{3+}$ ions in the samples. To the contrary, FTs of Mn K-edge spectra show a complicated structure related to an interference of different broadly spread Mn-O distances. As iron content increases, the FTs of Mn *K*-edge EXAFS spectra evolve and the multi-peak structure tend to converge into a single dominant peak at $x=0.7$. The increase in the first peak height seems to indicate the reduction of the degree of tetragonal distortion within $MnO_6$ octahedra. Crystallographic parameters for the room-temperature monoclinic structure of $Mn_2BO_4$ were adopted to calculate theoretical amplitude and phases for each scattering path up to 7 Å. The crystal structure of $Mn_2BO_4$ includes two types of $MnO_6$ octahedra: a strongly distorted $Mn^{3+}O_6$ and a more regular $Mn^{2+}O_6$ ones. The crystallographic structure of $Mn_2BO_4$ [8] was used as a starting model to fit the experimental spectra of doped $Mn_{2-x}Fe_xBO_4$ warwickites. The threshold energy shift $\Delta E_0$ and amplitude reduction factor $S_0^2$ were fixed at values 3 eV and 0.85 obtained for $Mn_2BO_4$. Six interatomic distances $R_{Mn-O}$ with a common Debye-Waller (DW) factor $\sigma^2$ were varied to obtain best fits. The same method was applied to fit the data at Fe *K*-edge with CoFeBO$_4$ as a reference compound [28]. The comparison between best fit and experimental spectra in the term of $k^3$-weighted FTs of EXAFS signals and back Fourier-transformed in $q$-space for $Mn_{1.7}Fe_{0.3}BO_4$ are shown in Figure SM1 [25]. The main results of the structural analysis at Fe and Mn $K$-edges are summarized in Table 2. The average interatomic distances $\langle$Fe-O$\rangle$ slightly decrease with decreasing Fe content. This can result from local strains in FeO$_6$ octahedra induced by neighboring Mn$^{3+}$O$_6$ octahedra with large tetragonal distortions. As Fe content increases, the contribution of Mn$^{2+}$O$_6$ to the average Mn-O distances increases that effectively leads to the monotonic increase in $\langle$Mn-O$\rangle$. The increase in the average $\langle$Mn-O$\rangle$ distance for $x$=0.7 agrees with the small chemical shift observed from XANES spectra. The tetragonal distortion of MnO$_6$ octahedral continuously reduces on going from $x$=0 to $x$=0.7, resulting in a decrease of the DW factors. This monotonic evolution is consistent with the single-ion JT mechanism.

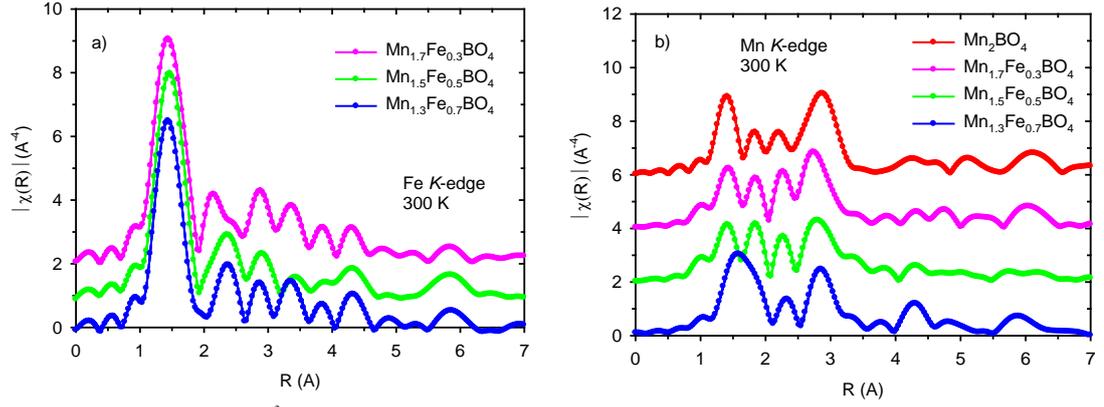

**Fig. 8.** FT modulus of the $k^3$-weighted EXAFS spectra of the $Mn_{2-x}Fe_xBO_4$ warwickites at the Fe (a) and Mn (b) $K$-edges at T=300 K. The curves have been vertically shifted for clarity.

**Table 2**
Best-fit structural parameters of the first oxygen coordination shell for $Mn_{2-x}Fe_xBO_4$ ($x$=0.0, 0.3, 0.5, 0.7) at the Mn and Fe $K$-edges. $N$ is the coordination number, $R$ is the interatomic distances for the octahedral site, $\sigma^2$ are the DW factors and $R_f$ is the fitting discrepancy factor. The average interatomic distance $\langle R_{M-O}\rangle$ is highlighted bold.

|   | $N$ | $R_{Mn-O}$ (Å) | $\sigma^2_{Mn-O}\cdot 10^{-3}$ (Å$^2$) | $R_{Mn-O}$ (%) | $R_{Fe-O}$ (Å) | $\sigma^2_{Fe-O}\cdot 10^{-3}$ (Å$^2$) | $R_{Fe-O}$ (%) |
|---|---|---|---|---|---|---|---|
| $Mn_2BO_4$ | 1 | 1.92(2) | | | | | |
| | 1 | 1.99(2) | | | | | |
| | 1 | 2.18(2) | | | | | |
| | 1 | 2.18(2) | 2.94 | 0.18 | | | |
| | 1 | 2.40(2) | | | | | |
| | 1 | 2.44(2) | | | | | |
| | | **2.19(2)** | | | | | |
| $Mn_{1.7}Fe_{0.3}BO_4$ | 1 | 1.96(2) | | | 1.79(2) | | |
| | 1 | 2.09(2) | | | 1.89(2) | | |
| | 1 | 2.22(2) | | | 1.89(2) | | |
| | 1 | 2.22(2) | 1.57 | 1.04 | 2.01(2) | 3.3 | 2.32 |
| | 1 | 2.37(2) | | | 2.01(2) | | |
| | 1 | 2.48(2) | | | 2.01(2) | | |
| | | **2.23(2)** | | | **1.93(2)** | | |
| $Mn_{1.5}Fe_{0.5}BO_4$ | 1 | 1.95(2) | | | 1.95(2) | | |
| | 1 | 2.08(2) | | | 1.95(2) | | |
| | 1 | 2.21(2) | | | 1.97(2) | | |
| | 1 | 2.21(2) | 0.6 | 4.5 | 1.97(2) | 8.7 | 2.9 |
| | 1 | 2.34(2) | | | 1.97(2) | | |
| | 1 | 2.46(2) | | | 2.13(2) | | |
| | | **2.21(2)** | | | **1.99(2)** | | |
| $Mn_{1.3}Fe_{0.7}BO_4$ | 1 | 2.06(2) | | | 1.93(2) | | |
| | 1 | 2.09(2) | | | 1.93(2) | | |
| | 1 | 2.23(2) | | | 1.97(2) | | |
| | 1 | 2.24(2) | 0.7 | 0.2 | 1.97(2) | 7.9 | 2.8 |
| | 1 | 2.31(2) | | | 2.03(2) | | |
| | 1 | 2.49(2) | | | 2.16(2) | | |
| | | **2.24(2)** | | | **1.99(2)** | | |

*3.4. Magnetization data*

A careful study of $Mn_{2-x}Fe_xBO_4$ magnetic properties is beyond the scope of this article and will be published elsewhere [29]. Here we present only preliminary results of magnetization measurements at low magnetic fields. Figure 9 shows the temperature dependences of *dc* magnetizations of $Mn_{2-x}Fe_xBO_4$ single crystals measured at a field of 250 Oe applied parallel to the crystallographic *c*-axis. Field cooled (FC) and zero field cooled (ZFC) magnetization curves start to diverge at temperatures below the maximum. The ZFC curves show a dropwise decrease of magnetization at extremely low temperatures, whereas FC curves tend to come to saturation as shown in inset to Fig. 9c for $Mn_{1.3}Fe_{0.7}BO_4$ as example. Critical temperatures of $T_{SG}$ = 11, 14 and 18 K are assigned to the spin-glass transition in $Mn_{2-x}Fe_xBO_4$ series with *x*=0.3, 0.5 and 0.7, respectively. A Curie-Weiss law is obeyed at highest temperatures with a negative paramagnetic temperature θ>|-170| K for all samples. As temperature decreases the deviation from the Curie-Weiss law arises. We have observed such a potential deviations in all the compounds in the $\chi^{-1}(T)$ curves that are attributed to the development of magnetic short-range order correlations.

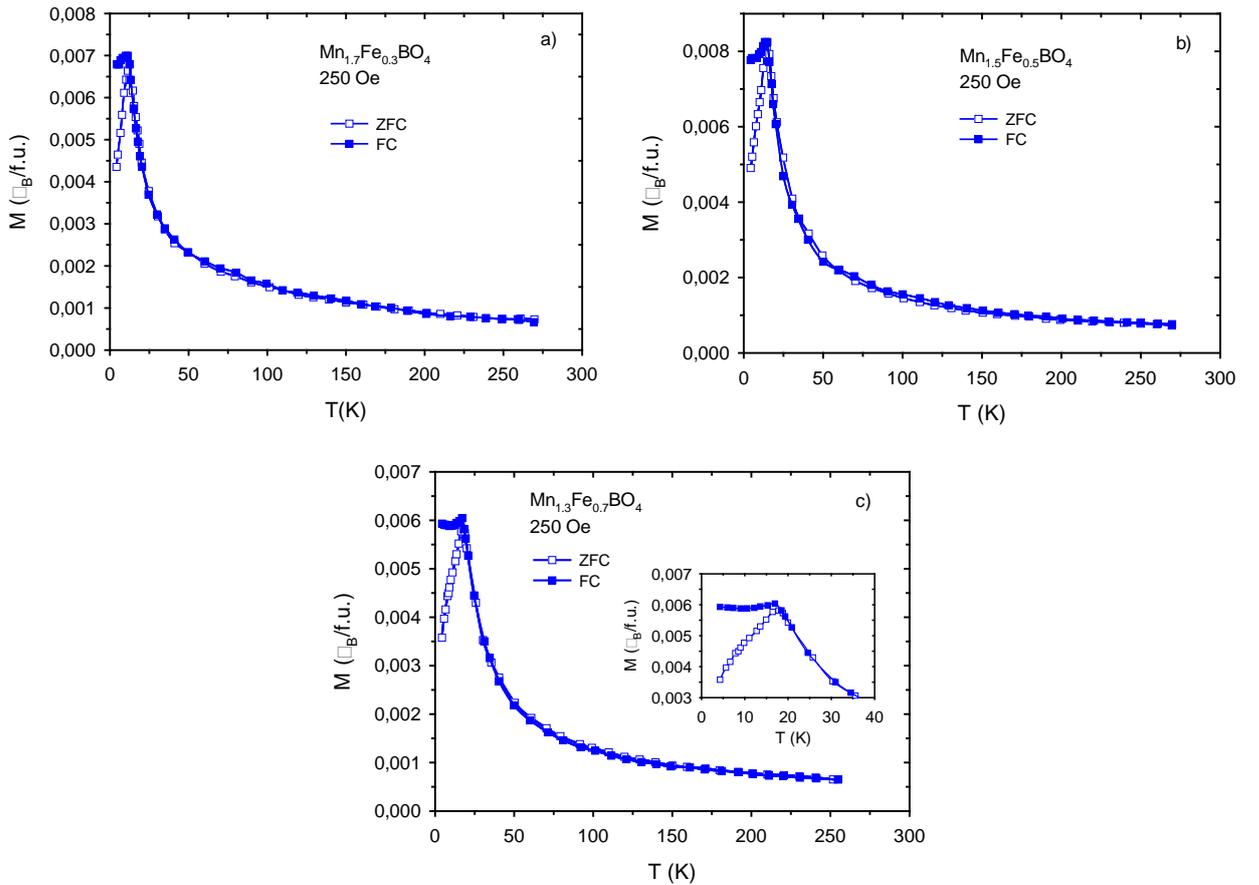

**Fig. 9.** Temperature dependence of magnetization of $Mn_{2-x}Fe_xBO_4$ samples at field H=250 Oe applied parallel to the crystallographic *c*-axis. The inset shows the splitting of ZFC and FC curves for $Mn_{1.7}Fe_{0.3}BO_4$.

## 4. Discussion

The synthesis of the compounds containing heterovalent manganese cations under the high process temperatures is complex problem due to the partial oxygen lose and the transition of the manganese oxide (3+) $Mn_2O_3$ to the manganese oxide (2+, 3+) $Mn_3O_4$ at the temperatures 950-1100°C. It makes some uncertainty of $Mn^{2+}$ and $Mn^{3+}$ cations concentration. In this work the fluxes were prepared at the temperature of 1100°C that coincides to the upper boundary of the oxide $Mn_2O_3$ decomposition temperature range. However the X-ray diffraction and XANES studies have showed that the composition *x* of the synthesized single crystals is agree with the initial ratio of the Mn-Fe in the fluxes. It could denote the hierarchy of the chemical bonds in the flux at the forming of the warwickite. It is possible that the iron with oxidation state 3+ has a priority at the crystal forming than the manganese 3+. The excess of the manganese become to be the part of the solvent.

The crystal structures of end member $Mn_2BO_4$ is described by the monoclinic (sp. gr. *P2₁/n*) syngony. The charge ordering of the type $Mn^{2+}(2)$-$Mn^{3+}(1)$ has been found to be associated with a strong Jahn-Teller distortion of the $Mn^{3+}O_6$ octahedra in the presence of orbital ordering ($d_z^2$). The Fe-substitution ($x \geq 0.3$) was found to promote a crystal structure with a higher symmetry (*Pnam*). The tetragonal distortion of $MnO_6$ octahedral continuously is reduced on going from $x=0$ to $x=0.7$, resulting in a decrease of the DW factors. This monotonic evolution is consistent with the single-ion JT mechanism. Shannon [30] listed the $Fe^{3+}$ and $Mn^{3+}$ as having the same ionic radius ($r_{Mn}^{3+} = r_{Fe}^{3+} = 0.645$ Å). However, it is apparent from Fig. 3 that the Fe addition causes the reduction of the unit cell parameters. The similar behavior was observed earlier for Mg-Co-Fe warwickite [28], and Mg-Co-Ga ludwigite [31] and was attributed to the high level of disorder in the substituted systems.

In the warwickite structure the magnetic ions inside the ribbon are built into a net of triangles with inter-ionic distances of ~3 Å (Fig. 2). In the presence of the antiferromagnetic exchange interactions between the nearest neighbors the geometric frustration of the exchange interactions are expected to appear. The inter-ribbon interactions become important at low temperatures, which leads either to the emergence of long-range magnetic order or freezing of spin dynamics. So, the spin-glass transitions have been found in the heterometallic $MgVBO_4$ (S=1) [1], $MgCrBO_4$ (S=3/2) [1], $MgFeBO_4$ (S=5/2) [1, 6, 32] and $CoFeBO_4$ ($S^{Co2+}=3/2$, $S^{Fe3+}=5/2$) [6, 14, 32] with monotonically increasing transition temperatures $T_{SG}$= 6, 6.5, 12 and 22 K respectively. The magnetization study has revealed the effect of Fe substitution on the magnetic properties of $Mn_{2-x}Fe_xBO_4$ warwickites. While $Mn_2BO_4$ shows long-range antiferromagnetic order, these compounds were found to exhibit a spin-glass transition below $T_{SG}$ =11, 14 and 18 K for $x=0.3$, 0.5 and 0.7, respectively. The large negative paramagnetic temperatures θ indicate the predominance of antiferromagnetic couplings. The Fe-induced cation disorder superimposed on the triangle network of the magnetic moments gives rise to enhance of the magnetic frustration role that is reflected in the large ratio $|\theta|/T_{SG} > 10$.

## 5. Conclusion

In the present study the flux based on the bismuth trimolybdate, diluted by the sodium carbonate, was used for synthesis of the $Mn_{2-x}Fe_xBO_4$ warwickites. Due to the much lower viscosity of this flux than the $B_2O_3$ fluxes, it was possible to obtain single-crystalline samples of suffitient size. Single crystals have been studied by means of X-ray absorption spectroscopy at both Mn and Fe *K*-edges. The substitution of $Mn^{3+}$ by $Fe^{3+}$ has been found and a weak mixing of the $Mn^{2+}$ and $Mn^{3+}$ electronic states have been confirmed by XANES spectroscopy. The local geometric structure around $Fe^{3+}$ slightly changes along the series reflecting the decrease in the local strains within $FeO_6$ octahedra induced by highly distorted $Mn^{3+}O_6$ octahedra. The local structure around Mn atoms is modified more seriously. The changes can be generally described as a decrease in the degree of tetragonal distortions within the $MnO_6$ octahedra. The Fe-substitution induces the crystal symmetry change from monoclinic ($x=0$) to the orthorhombic ($x\geq0.3$) one. The charge ordering typical of $Mn_2BO_4$ is broken and the random distribution of Mn and Fe ions over the metal sites appears. The Fe-doping enhances the magnetic frustation giving rise to the supression of the long-range magnetic order (AF vs. spin-glass in $Mn_2BO_4$ and Fe-substituted warwickites, respectively).

## Acknowledgments
This work has been financed in part by the grants of the Council for Grants of the President of the Russian Federation (project nos. SP-938.2015.5, NSh-7559.2016.2), the Russian Foundation for Basic Research (project nos. 16-32-60049 mol_a_dk, 16-32-00206 mol_a, 17-02-00826-a, 16-32-50058).